\pdfoutput=1
\RequirePackage{amsmath}
\documentclass[runningheads]{llncs}
\usepackage[T1]{fontenc}
\usepackage{graphicx}
\usepackage{pifont}
\usepackage[acronym]{glossaries}
\usepackage{multirow}
\usepackage{svg}
\usepackage{orcidlink}
\usepackage{hyperref} 
\hypersetup{hidelinks}  
\usepackage{color}

\newcommand{\cmark}{\ding{51}}%
\newcommand{\xmark}{\ding{55}}%

\def\preprint{1} 

\if\preprint1
\usepackage[all]{background}
\usepackage{stackengine}
\fi

\begin{document}
\title{parti-gem5: gem5's Timing Mode Parallelised}
%
%

\if\preprint1
\setstackEOL{\\}
\setstackgap{L}{\normalbaselineskip}
\SetBgContents{\color{gray}{\tiny \Longstack{PREPRINT - Accepted at Embedded Computer Systems: Architectures, Modeling, and Simulation \\
                                             (SAMOS) Conference 2023}}}
\SetBgPosition{4.5cm,1cm}
\SetBgOpacity{1.0}
\SetBgAngle{0}
\SetBgScale{1.8}
\fi

\author{José Cubero-Cascante \orcidlink{0000-0001-9575-0856} \and
Niko Zurstraßen \orcidlink{0000-0003-3434-2271} \and
Jörn Nöller \orcidlink{0000-0001-6115-065X}  \and \\
Rainer Leupers \orcidlink{0000-0002-6735-3033} \and
Jan Moritz Joseph \orcidlink{0000-0001-8669-1225}}
\authorrunning{J. Cubero-Cascante et al.}
\institute{Institute for Communication Technologies and Embedded Systems,\\
RWTH Aachen University, Aachen, Germany
\email{\{cubero,zurstrassen,noeller,leupers,joseph\}@ice.rwth-aachen.com}}

\maketitle
%
\newacronym[plural=ICs,firstplural=Integrated Circuts (ICs)]{ic}{IC}{Integrated Circuit}
\newacronym[plural=MPSoCs,firstplural=Multiprocessor Systems on Chip (MPSoCs)]{mpsoc}{MPSoC}{Multiprocessor System on Chip}
\newacronym[plural=FSSs,firstplural=Full System Simulators (FSSs)]{fss}{FSS}{Full System Simulator}
\newacronym{pdes}{PDES}{Parallel Discrete Event Simulation}
\newacronym{des}{DES}{Discrete Event Simulation}
\newacronym{esl}{ESL}{Electronic System Level}
\newacronym{nic}{NIC}{Network Interface Controller}
\newacronym{ksloc}{kSLOC}{kilo Software Lines Of Code}
\newacronym{npb}{NPB}{NAS Parallel Benchmarks}
\newacronym{kvm}{KVM}{Kernel Virtual Mode}
\newacronym[plural=KPIs,firstplural=Key Performance Indicators (KPIs)]{kpi}{KPI}{Key Performance Indicator}
\newacronym[plural=ISAs,firstplural=Instruction Set Architectures (ISAs)]{isa}{ISA}{Instruction Set Architecture}
\newacronym[plural=TLBs,firstplural=Transaction Lookaside Buffers (TLBs)]{tlb}{TLB}{Transaction Lookaside Buffer}
\newacronym[plural=ROIs,firstplural=Regions of Interest (ROIs)]{roi}{ROI}{Region of Interest}
\newacronym{ipc}{IPC}{Instructions Per Second}
\newacronym{pci}{PCI}{Peripheral Component Interconnect}
\newacronym{dma}{DMA}{Direct Memory Access}
\newacronym{slicc}{SLICC}{Specification Language for Implementing Cache Coherence}
\newacronym{mips}{MIPS}{Million Instructions Per Second}
\newacronym{noc}{NoC}{Network on Chip}
\newacronym[plural=UARTs,firstplural=Universal Asynchronous Receiver-Transmitters (UARTs)]{uart}{UART}{Universal Asynchronous Receiver-Transmitter}

\begin{abstract}
Detailed timing models are indispensable tools for the design space exploration of \glsfirstplural{mpsoc}.
As core counts continue to increase, the complexity in memory hierarchies and interconnect topologies is also growing,
making accurate predictions of design decisions more challenging than ever.
In this context, the open-source \glsfirst{fss} gem5 is a popular choice for MPSoC design space exploration,
thanks to its flexibility and robust set of detailed timing models.
However, its single-threaded simulation kernel severely hampers its throughput.
To address this challenge, we introduce \emph{parti-gem5}, an extension of gem5 that enables parallel timing simulations on modern multi-core simulation hosts.
Unlike previous works, \emph{parti-gem5} supports gem5's timing mode, the \texttt{O3CPU}, and Ruby's custom cache and interconnect models.
Compared to reference single-thread simulations, we achieved speedups of up to 42.7$\times$ when simulating a 120-core ARM \glsname{mpsoc} on a 64-core x86-64 host system.
While our method introduces timing deviations, the error in total simulated time is below 15\% in most cases.
\end{abstract}

\section{Introduction}

\glspl{mpsoc} are present in a broad range of platforms, from web servers to automotive control units and smartphones.
In addition to many cores, these systems include complex memory hierarchies with multiple levels of coherent caches and custom interconnects.
System architects require detailed timing models to study the impact of hardware design choices on the overall system performance.
\glspl{fss} are a powerful tool to accomplish this task.

Among these, gem5~\cite{gem5} is an open-source system-level simulator widely used in industry and academia.
It has been actively maintained and developed for more than thirty years~\cite{gem52020} and
supports all major \glspl{isa} as targets, including ARM, x86 and RISC-V.
gem5 is based on a \gls{des} and includes detailed models for CPU cores, caches, memory, and interconnect.

For modelling memory transactions, two main simulation modes are available.
The atomic mode supports modelling functional behaviour only, and its usage is limited to software tests or quickly advancing a simulation to \glspl{roi}.
The timing mode covers both functional and timing behaviour.
With the timing mode, it is possible to study the performance impact of the interconnect, such as queuing delays or resource contention.
Optimisations like pipelining and out-of-order execution can also be modelled.
Hence, only the timing mode can be employed for micro-architecture exploration.

One significant problem of gem5 is its single-threaded \gls{des} kernel, which limits the performance and scalability of the simulations.
This issue is even more severe when the target is a multi-core system.
Based on our benchmarks, the timing mode achieves between 0.01 and 0.1 \gls{mips} on a recent high-performance workstation.
If the target system is a modern \gls{mpsoc} capable of achieving 64,000 \gls{mips}, one second of the target's time would require between 1 and 10 weeks of host time.

A previous contribution, \emph{par-gem5}~\cite{pargem5}, extended gem5 to support parallel simulations
but its usage is limited to the atomic mode.

Table~\ref{tab:cpumodels} summarises the capabilities of gem5's main CPU models, their timing detail level, and their customisation support.
Their work enables parallel simulation for the detailed timing CPU models, which were not tackled by any other work so far.
With \emph{parti-gem5}, we leverage full parallelisation for every use case, from checkpointing to design space exploration.

Our main contributions are:
\begin{itemize}
  \item We extend par-gem5 by supporting the detailed \texttt{MinorCPU} and \texttt{O3CPU} models.
  \item We present a thread-safe message passing mechanism that enables the modelling of coherent \glspl{mpsoc} using the Ruby~\cite{gems_ruby} memory system.
  \item We demonstrate speedups of up to 42.7$\times$ while keeping the error in simulation statistics below 15\% in most cases.
\end{itemize}

\begin{table}
  \centering
  \caption{Main CPU Models in gem5 and their Timing Features.}
  \begin{tabular}{|c|c|c|c|c|}
    \hline
    \textbf{CPU model} & \textbf{KVM} & \textbf{Atomic} & \textbf{Minor} & \textbf{O3}\\
    \hline
    Pipeline & N/A & none & in-order & out-of-order \\
    \hline
    Communication protocol & N/A & atomic & timing & timing \\
    \hline
    Custom cache protocols (Ruby)& \xmark & \xmark & \cmark & \cmark \\
    \hline
    Custom interconnect (Ruby)& \xmark & \xmark & \cmark & \cmark \\
    \hline
    Parallel simulation & gem5~\cite{gem5,gem52020} & par-gem5~\cite{pargem5} & our work & our work \\
    \hline
  \end{tabular}
  \label{tab:cpumodels}
\end{table}

The rest of this paper is organised as follows: A summary of related work is presented in Section~\ref{sec:related}.
Section~\ref{sec:background} explains the key concepts pertaining to gem5 timing simulations.
Section~\ref{sec:methods} presents the challenges for parallelising timing simulations and our approach to tackling them.
We show our experimental evaluation of \emph{parti-gem5} regarding performance and accuracy in Section~\ref{sec:results}.
Finally, we conclude in Section \ref{sec:conclusion}.

\section{Related Work}\label{sec:related}

\gls{pdes} and its application to \glspl{fss} have been active areas of research for several decades
with fundamental theoretical work laid in~\cite{pdes_chandy_1979,pdes_fujimoto_1990}.
When distributing a simulation among several threads, consistently advancing time becomes a major challenge.
In this regard, simulators can be either classified as synchronous or asynchronous~\cite{parsc}.

In asynchronous simulators, each thread keeps track of the times of all other threads and is allowed to advance its own time if it stays within a maximum look-ahead distance $t_{la}$ to the slowest thread.
This approach is used in \emph{Manifold}~\cite{manifold_2014} and \emph{SCope}~\cite{weinstock2016_journal}.

By contrast, synchronous simulators use global synchronisation mechanisms.
Among these, delta-cycle simulators, such as \cite{parsc}, enforce synchronisation at every single timestamp.
However, limited performance gains are observed in typical \gls{fss} scenarios as the amount of simultaneous activity is often not high enough to overcome the synchronisation overhead\cite{weinstock2016_journal}.
Quantum-based protocols~\cite{wwt93,sniper2011,cos2020,distgem5} are another type of synchronous simulator.
The simulation time is divided into windows of length $t_{q\Delta}$, called quanta, in which threads can advance independently.
Synchronisation is only performed at global barrier events, leading to higher speedups.

In asynchronous and window-based synchronous approaches, each thread has a local time.
This poses the risk of causality errors as there is no guarantee of a chronological execution order for the events.
To prevent these errors, most PDES implementations fall back to either conservative or optimistic methods~\cite{pdes_fujimoto_1990}.
With optimistic methods, such as ~\cite{spectempfork_2019}, causality errors are not prevented in the first place but are detected and corrected subsequently.
Conservative methods, as in ~\cite{distgem5,cos2020} prevent causality errors by imposing constraints on the models' behaviour.
However, some simulators~\cite{manifold_2014,weinstock2016_journal} employ domain-specific timing information and trade-off minor timing deviations against simulation speed.

In the scope of gem5, some efforts have been made to enable parallel simulations.
Parallel support for KVM was integrated in 2013 to mainstream gem5~\cite{gem52020}.
The authors from~\cite{gem5kvm} introduce a parallel execution mode called \emph{pFSA} which exploits \gls{kvm}~\cite{kvm} to fast-forward gem5 simulations to \glspl{roi} at almost native speeds.
\gls{pdes} is not used to simulate the \glspl{roi}, but several independent detailed simulations are run in parallel.
One limitation of KVM is that it can only be used when the target and host platform match.
\emph{dist-gem5}~\cite{distgem5} and \emph{COSSIM}~\cite{cos2020} are gem5 extensions targetting distributed systems.
For instance, in \emph{dist-gem5}, multiple compute nodes are assumed to communicate via \glspl{nic} with a known latency.
A fast and perfectly accurate simulation is attained by setting $t_{q\Delta}$ smaller or equal to the \gls{nic} latency.
This, however, limits the applicability of \emph{dist-gem5} to the simulation of distributed systems connected via \glspl{nic}.

The more general \emph{par-gem5}~\cite{pargem5} allows to simulate \glspl{mpsoc}.
Their quantum-based approach allows causality errors but minimises their occurrence with a carefully-chosen system partitioning and quantum setting.
This extension yields a significant acceleration of gem5 full system simulations, with up to 24.7$\times$ speedups reported.
However, it is restricted to gem5's atomic mode, which has minimal timing details.

With \emph{parti-gem5}, we extend the work from \emph{par-gem5} by enabling the parallel simulation of \glspl{mpsoc} with the most detailed timing models:
the in-order \texttt{MinorCPU}, the out-of-order \texttt{O3CPU} and the Ruby coherent memory sub-system.

\begin{figure}
  \centering
  \includegraphics[width=0.8\linewidth]{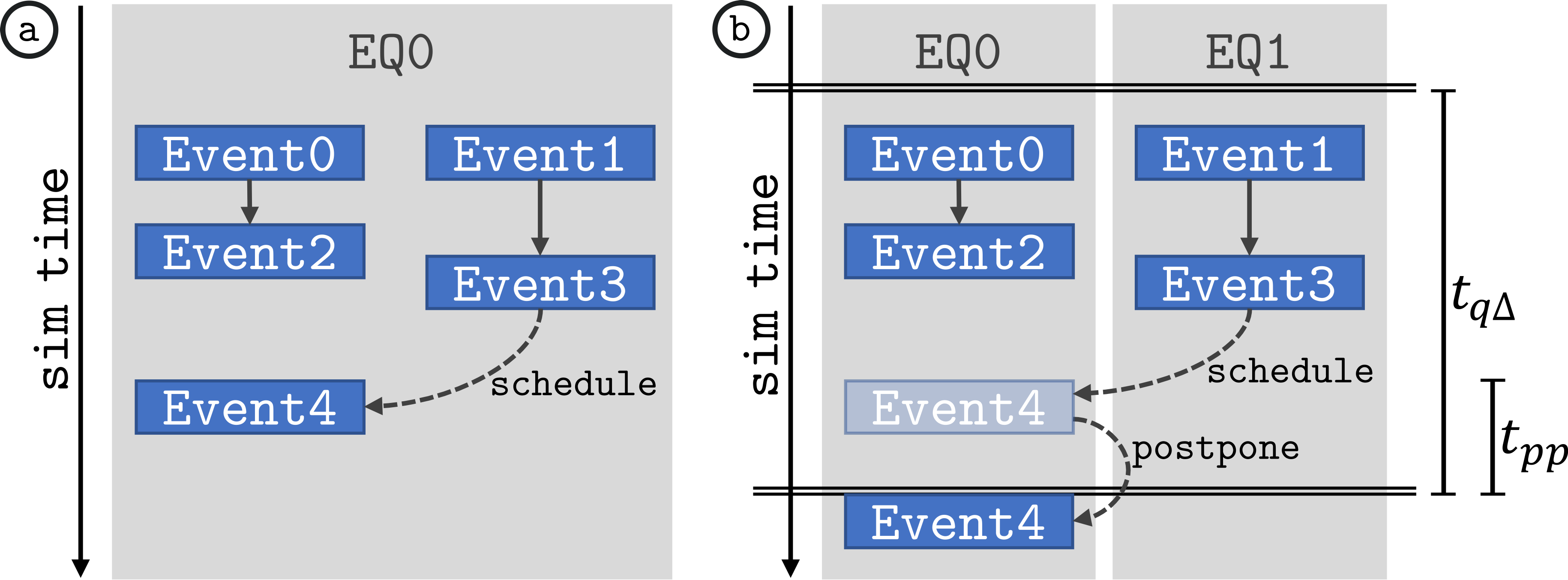}
  \caption{Flow diagram of gem5's DES and PDES. Adapted from~\cite{pargem5}.}
  \label{fig:des_pdes}
\end{figure}

\section{Background}\label{sec:background}

\subsection{Discrete Event Simulation in gem5}\label{sec:pdes}

The kernel of gem5 is a \gls{des} engine.
HW components are termed \texttt{SimObject}s, and their behaviour is modelled through events.
For an event to be executed, it must first be placed in an event queue (\texttt{EQ}) by calling \texttt{schedule()}.
Scheduled events are ordered based on the target time and priority.
A simulation thread processes all events in an event queue, one at a time, starting at its head.
When processed, an event can schedule, deschedule and reschedule new events.
In the default single-thread \gls{des} engine, only one event queue and one simulation thread are used.
This is depicted in Fig.~\ref{fig:des_pdes}a.

Synchronous de-coupled parallel simulation, as used in \emph{par-gem5}~\cite{pargem5}, is enabled by distributing the hardware objects into $N$ time domains.
Each domain has an independent event queue and processing thread.
The simulation time is divided into slices of fixed length $t_{q\Delta}$, called quanta, in which the event queues run independently and in parallel.
Barrier events at each quantum border ensure synchronisation between threads.
This \gls{pdes} approach is shown in Fig.~\ref{fig:des_pdes}b.

As depicted, an event may attempt to schedule another which belongs to a different queue.
Special handling is required in such cases as the exact current time of the target domain is unknown, and scheduling events in the past is not allowed.
We refer to this situation as inter-domain scheduling.
If the target schedule time is earlier than the next quantum barrier, the event is postponed to the next quantum border.
This introduces a delay $t_{pp} \in[0,t_{q\Delta}]$, which is an artefact of the parallelisation.
As in \cite{weinstock2016_journal}, such timing deviations are allowed, sacrificing accuracy for speedup.
However, their impact is minimised by carefully setting the quantum length and the boundaries of the simulation domains.

\subsection{CPU Models in gem5}\label{subsec:cpumodels}

Table~\ref{tab:cpumodels} shows the main CPU models available in gem5, which offer distinct levels of timing detail.
The \texttt{KVMCPU} executes software workloads natively on the host system, based on Linux's \gls{kvm}~\cite{kvm}.
Consequently, this model can attain near-native execution speeds but
does not generate any accurate statistics of the simulated platform.
It should only be used to fast-forward to \glspl{roi}.

The \texttt{AtomicCPU} has an interpreter-like core, executing instruction after instruction with a fixed delay.
It can be used to study basic cache behaviour or for software verification.

The in-order \texttt{MinorCPU} and out-of-order \texttt{O3CPU} provide the most detailed simulation.
These models include micro-architectural features like pipelining and memory access re-ordering.
The authors in~\cite{gem5_arm_cortex} show that these models provide excellent accuracy compared to real ARM \glspl{mpsoc}.
In their experiments, the error in total execution time lies between 1\% to 17\% for the \texttt{MinorCPU} and between 2\% to 16\% for the \texttt{03CPU}
with average errors of 7.4\% and 8\% respectively.
Furthermore, these models can be combined with the highly detailed Ruby Cache system, which enables accurate modelling of
cache coherence protocols and custom interconnects.

\subsection{Communication Protocols in gem5}\label{subsec:communication}

In gem5, SimObjects communicate via ports using packets, structures containing a target address, payload data, header and packet delay.
The way packets are transmitted is determined by the selected communication protocol.
Hereby, gem5 offers three protocols: functional, atomic and timing.
The functional protocol is used only for debugging purposes, so we refrain from further explanation.

\begin{figure}[bt]
  \centering
  \includegraphics[width=\textwidth]{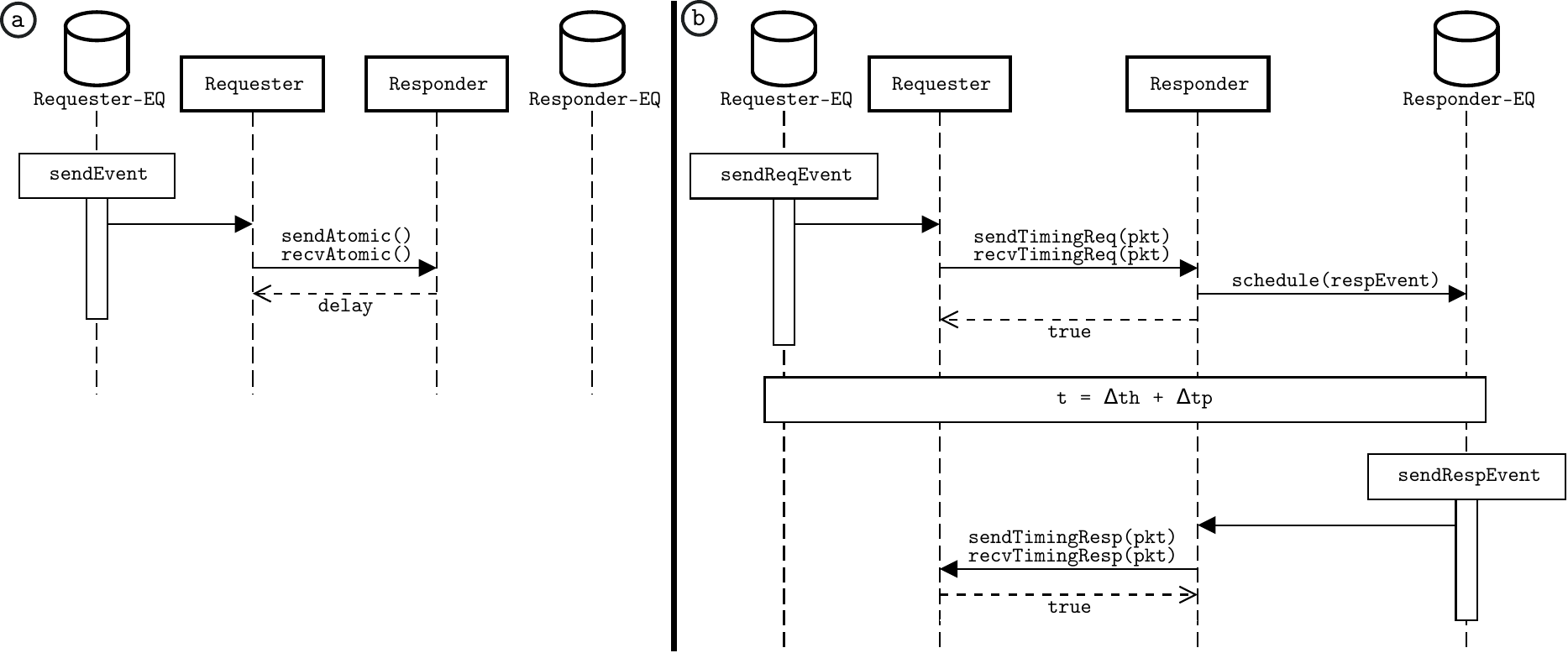}
  \caption{Communication between a requester and a responder using the Atomic Mode (a) and the Timing Mode (b).}
  \label{fig:timing_protocol}
\end{figure}

The atomic protocol is depicted in Fig.~\ref{fig:timing_protocol}a.
Transmitting a packet is achieved in one single event.
The requester only needs to call \texttt{sendAtomic(pkt0)},
which is received as \texttt{recvAtomic(pkt)} at the responder's side.
All effects that might arise from this transaction, including coherence snoops, are handled in this single call chain.

The timing protocol divides a transaction into two phases, implemented in separate events, as shown in Fig.~\ref{fig:timing_protocol}b.
The requester starts the transaction by calling \texttt{sendTimingReq(pkt)},
which is received as a \texttt{recvTimingReq(pkt)} at the responder's side.
If the responder accepts the request, it schedules a response event and returns \texttt{true}.
During the response event, the responder calls \texttt{sendTimingResp(pkt)},
which is received as \texttt{recvTimingResp(pkt)} at the requester's side.
Finally, the requester accepts the response by returning \texttt{true}.
Between the request and response events, the simulation time usually advances by $\Delta t_h + \Delta t_p$,
with $\Delta t_h$ being the header delay and $\Delta t_p$ the delay of the packet.
If a requester or responder is busy receiving a transaction, it can reject the packet by returning \texttt{false}.
The rejecter is responsible for signalling a retry once it is free again.
In such cases, more than two simulation events per transaction are required.

Due to this higher complexity, the timing protocol has a lower performance.
In our experiments, simulations using the timing protocol and the detailed \texttt{O3CPU}
yield only 20\% of the performance obtained with the atomic protocol and the \texttt{AtomicCPU}.

\subsection{The Ruby Cache and Interconnect Subsystem}\label{subsec:ruby}

Ruby, developed initially as part of the GEMS project~\cite{gems_ruby}, extends gem5 with highly configurable cache and interconnect models.
Cache transactions are simulated with high fidelity~\cite{gem52020}.
Many network topologies and cache coherence protocols are provided, and the user can define new ones using the \gls{slicc}.
Support for the ARM AMBA CHI protocol~\cite{arm_amba_chi}, widely used in modern high-performance ARM-based platforms, is available.

\begin{figure}[bt]
  \centering
  \includegraphics[width=0.8\textwidth]{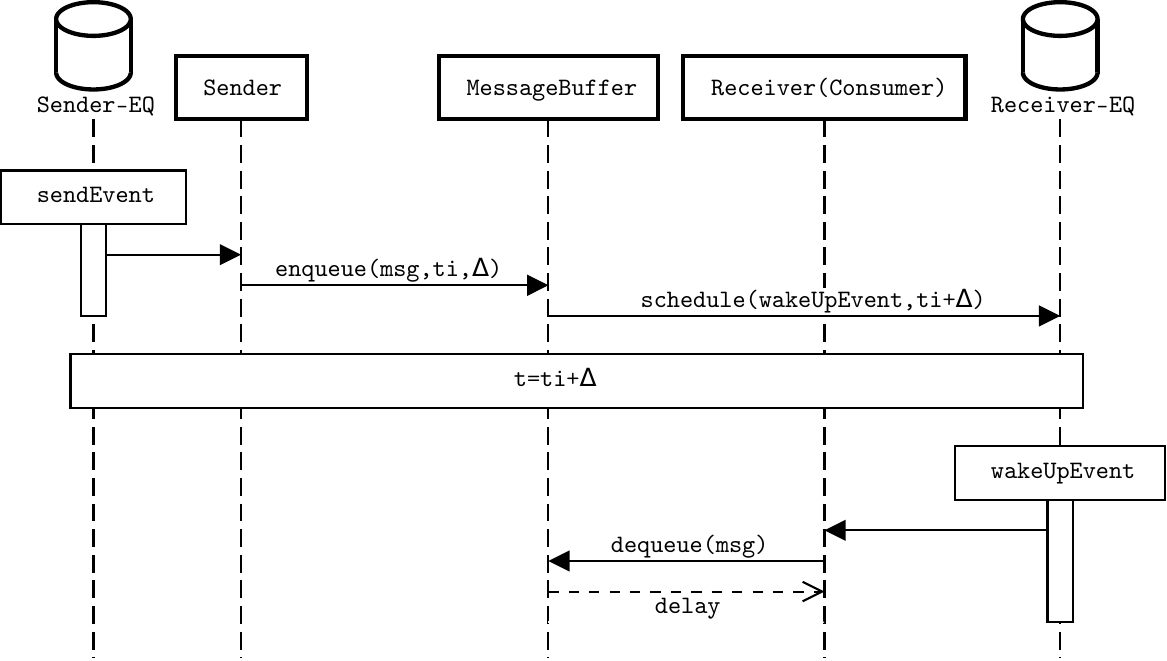}
  \caption{Ruby Message Passing}
  \label{fig:ruby-protocol}
\end{figure}

The Ruby subsystem consists on a set of interconnected nodes which communicate using a buffered message-passing protocol.
Sending a message between two Ruby nodes involves three main objects: the sender node, a \texttt{MessageBuffer} and the receiver node or \texttt{Consumer}.
The sender node represents the Ruby object that initiates or forwards the message.
The \texttt{MessageBuffer} models the communication link;
it contains a priority queue structure, which holds all messages in transit, ordered according to their insertion time.
The receiver node implements the abstract class \texttt{Consumer}.
Specialisations of this class, such as cache controllers and network routers, implement their specific behaviour through the virtual method: \texttt{wakeup()}.
Several message buffers can be assigned to the same \texttt{Consumer}, forming an $N:1$ relationship.

Fig.~\ref{fig:ruby-protocol} shows the relation between these three objects.
Transmitting a message from the sender to the \texttt{Consumer} requires two phases corresponding to separate simulation events.
In the first event, the sender node enqueues a new message in the \texttt{MessageBuffer}, providing a timing annotation \texttt{delta}.
This corresponds to the difference between the current time and the arrival time for the message.
The \texttt{enqueue()} method leads to the (re-)scheduling of a wakeup event on the assigned \texttt{Consumer} object.
During the wakeup event, the \texttt{Consumer} receives the message using the \texttt{dequeue()} function.
All buffers assigned to this \texttt{Consumer} will be checked in search for messages ready to be received at that point in time.

One weakness of Ruby stems from its independent development.
Most models, including the main transaction initiators, e.g. CPUs, GPUs and PCI modules, lack native support for the Ruby protocol.
The same is true for the primary transaction targets such as the DDR controller.
Hence, simulating full systems with Ruby requires constant conversions between timing protocol packets and Ruby messages.
This task is performed by the \texttt{Sequencer} modules.

A typical multi-core system modelled with Ruby is depicted in Fig.~\ref{fig:ruby-topo}.
Ruby nodes are used to model the caches and the coherent transactions, while the timing protocol is used for non-coherent transactions.
Non-coherent targets, like low-speed system peripherals and timers, are made available to the CPUs through a single input/output crossbar (\emph{IO-XBAR}).
In Fig.~\ref{fig:ruby-topo}, the colour of each module indicates their supported protocol:
Black modules like the CPU and the \emph{IO-XBAR} use the gem5 timing protocol, while blue objects like the interconnect routers use the Ruby protocol.
\section{Parallelising gem5 Timing Models}\label{sec:methods}

\subsection{System Partitioning and Main Parallelisation Challenges}\label{subsec:methods_partitioning}

As explained in Section~\ref{sec:pdes}, simulating in parallel requires partitioning the target system into time domains.
All objects in each domain share one simulation thread and one event queue.
Objects within one domain have their send and receive events handled by the same thread, so their timing remains unaltered.
On the contrary, sending packets across domains requires inter-domain scheduling and timing deviations might occur.
Hence, the main goal of partitioning is to minimise the number of data transactions crossing the border between event queue domains.

Our target system is arranged in a hierarchical pattern, following the modelled \gls{mpsoc} structure.
Following the approach from ~\cite{wwt93,sniper2011,pargem5},
we leverage this pre-defined hierarchy to define the time domains.
Each CPU core is allocated to one distinct domain.
Exclusive resources such as private caches, \glspl{tlb} and local interconnect are placed in their CPU's domain.
This way, transactions hitting the private caches are completed within the same domain, allowing cores to advance quickly in their execution.
Shared resources, such as global interconnect, global caches, main DDR memory and system peripherals, are placed in one additional domain.
With this approach, $N+1$ threads are needed for a total of $N$ simulated CPUs.

\begin{figure}[bt]
  \centering
  \includegraphics[width=\textwidth]{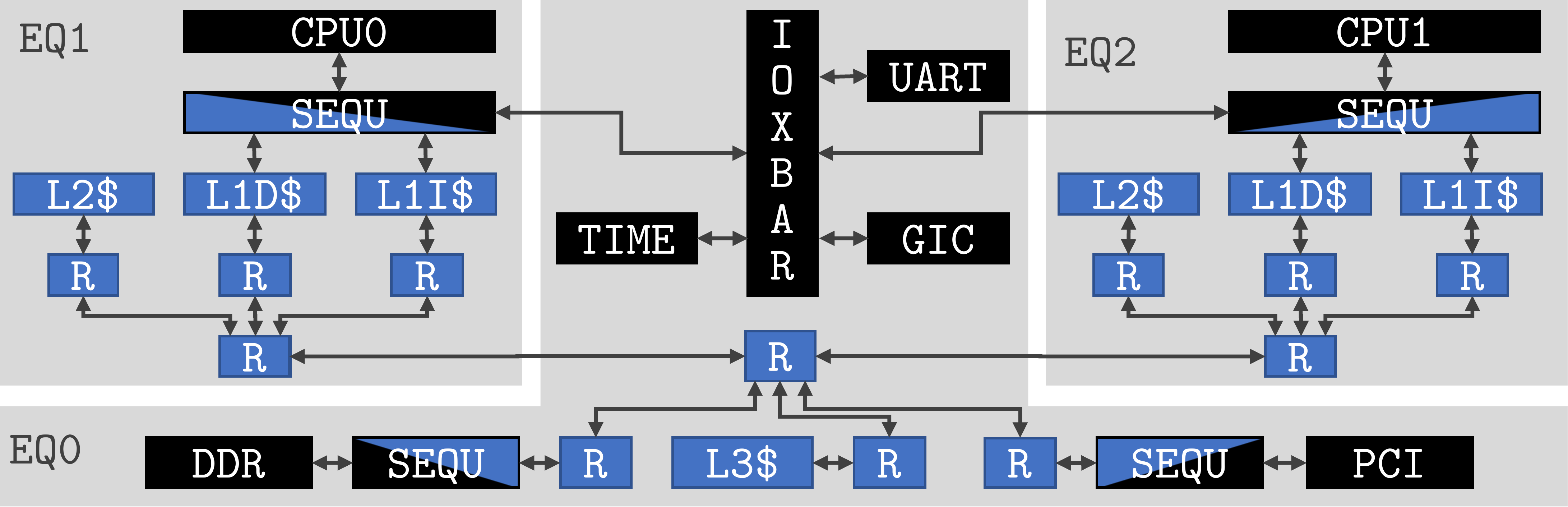}
  \caption{Topology and event queue assignment of an exemplary Ruby system.
  For the sake of simplicity, some components, like message buffers or TLBs are not depicted.}
  \label{fig:ruby-topo}
\end{figure}

Fig.~\ref{fig:ruby-topo} shows how a two-core system is partitioned into three time domains.
Two interconnect links cross the border between each CPU domain and the shared domain \texttt{EQ0}.
The link between local and central router objects employs the Ruby protocol,
while the link between the CPU's sequencer and the common \emph{IO-XBAR} uses the timing protocol.
Ensuring the correct behaviour for transactions passing through these links poses a significant challenge.
The following sections present the concepts applied at these two critical interfaces to enable a thread-safe parallel simulation.

\subsection{Thread-safe Ruby Message Passing}

The Ruby protocol has inherent properties that can be leveraged for parallel execution.
It follows a producer-consumer pattern in which the message buffer is the only shared structure.
After partitioning the system, some Ruby links have sender and receiver nodes residing in different domains.
Thread safety requires protecting the buffers with a mutual exclusion strategy.
However, as explained next, using an independent mutex to protect each buffer is insufficient to ensure proper communication between Ruby nodes.

\begin{figure}[bt]
  \centering
  \includegraphics[width=\textwidth]{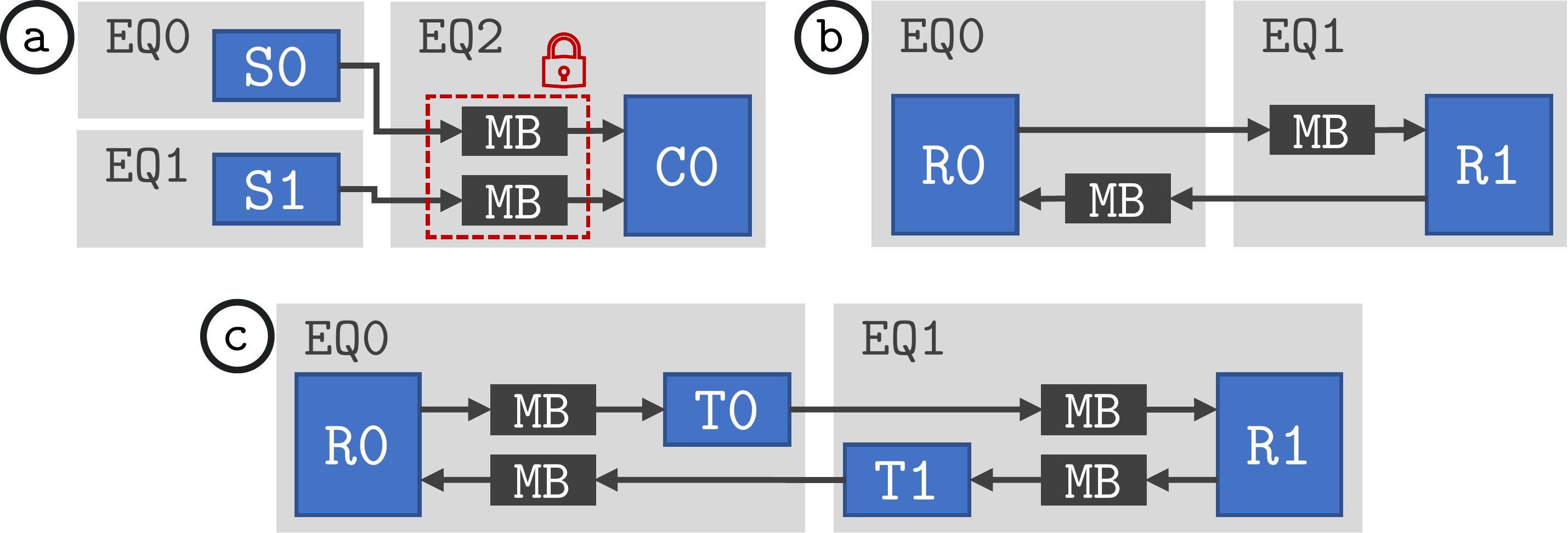}
  \caption{Ruby parallelisation challenges and solutions. (a) Multiple senders \texttt{S0} and \texttt{S1} communicate with a single consumer \texttt{C0}.
  (b,c) Bi-directional message passing between two routers \texttt{R0} and \texttt{R1}.
  The circular wait in (b) is eliminated in (c) by introducing the \texttt{Throttle} objects \texttt{T0} and \texttt{T1}.}
  \label{fig:ruby_mutexes}
\end{figure}

Some consumer objects receive messages from several senders, which might belong to several domains in a parallel simulation.
This situation is depicted in Fig.~\ref{fig:ruby_mutexes}a, where senders \texttt{S0} and \texttt{S1} have \texttt{C0} as common consumer.
In some cases, sender objects check the state of their output buffers before inserting a new message.
For example, the sender might use the number of available slots to decide if the link is available.
During the wakeup event, the consumer object checks all its input buffers, one at a time, and triggers the necessary actions to handle the incoming messages.
Hence, it is necessary to guarantee a serial execution of the consumer wakeup event and the events from all linked senders.

We realise this with a shared mutex concept, illustrated in Fig.~\ref{fig:ruby_mutexes}a.
Each consumer gets a unique wakeup mutex object during initialisation.
All input message buffers assigned to this consumer share this same mutex instance, depicted as a single red dash bounding box in Fig.~\ref{fig:ruby_mutexes}a.
The consumer locks the mutex during its wakeup event, preventing the arrival of new messages to any of its input buffers.
When done, the consumer releases the wakeup mutex, allowing the senders to enqueue new messages serially.

Another situation to be tackled is bi-directional communication.
Routers, for example, act both as sender and consumer and must be able to communicate with each other.
Such a topology could be implemented by using one message buffer for the forward link and another for the backward link, as shown in Fig.~\ref{fig:ruby_mutexes}b.
However, this configuration creates a circular wait which might result in a deadlock, a problem discussed in \cite{pdes_fujimoto_1990}.
If the wakeup events of R0 and R1 occur simultaneously, both routers will lock their input buffers and wait indefinitely for the availability of their output buffers.

A deadlock-free implementation is created by adding one object and one message buffer to the outputs of each router. See Fig.~\ref{fig:ruby_mutexes}c.
With this arrangement, the domain border is crossed by two independent uni-directional links: \texttt{T0-R1} and \texttt{T1-R0}.
The objects \texttt{T0} and \texttt{T1} represent instances of the Ruby class \texttt{Throttle}.
Throttle objects, placed at the output of each router, are used to model the link's bandwidth.
Our parallelisation approach profits from this separation of concerns between the router and the throttle classes.

\subsection{Thread-safe Concurrent Non-coherent Traffic}
\begin{figure}[bt]
  \centering
  \includegraphics[width=0.5\textwidth]{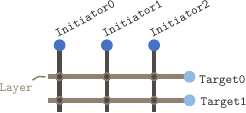}
  \caption{Example of a 3x2 non-coherent IO-Crossbar in gem5.}
  \label{fig:ioxbar}
\end{figure}

The non-coherent Input-Output Crossbar \emph{IO-XBar} is an N-to-M network node allowing the CPU cores to send transactions to peripheral devices, such as \glspl{uart} or timers.
Concurrent transactions involving two disjoint sender-receiver pairs are possible.
However, concurrent transactions from two different senders to the same target must be serialised.
This is accomplished using the concept of layers.
A layer is a communication channel to one target and can only be occupied by one initiator at a time (see Fig.~\ref{fig:ioxbar}).
Whenever an initiator wants to send a message to a target, it first tries to occupy the corresponding layer.
If another initiator already holds the layer, further requests are rejected.
Once an initiator has claimed the layer, it can communicate with the target using the gem5 timing protocol.
Every initiator bears the responsibility to release the layer after a communication.
This release is accomplished by scheduling a so-called release event, eventually informing rejected initiators to retry their transmissions.

In \emph{parti-gem5}, multiple CPUs can compete for a layer at the same host time, even if their local simulated times differ.
From the simulation time perspective, these are not concurrent accesses.
However, when not adequately handled, race conditions arise.
Our solution treats this situation as a special case of concurrent transactions.
We extend the layer concept by protecting the crossbar layer's state with mutexes and rejecting incoming transactions if the mutex has already been locked.
This slight adaptation introduces thread safety to the existing occupy and the retry mechanisms.

\section{Experimental Evaluation}\label{sec:results}

The main objective of our experiments is to evaluate the performance and accuracy of \emph{parti-gem5} compared to the standard single-thread gem5.
We measure performance as speedup, the ratio between the host execution times of a reference single-thread simulation and the evaluated parallel simulation.
For accuracy, we show the simulation error as the percentual deviation in total simulated time.
This is a good indicator of the overall accuracy since individual timing deviations introduced by the parallelisation will ultimately be reflected there.
As a second accuracy indicator, we also show the absolute error on the miss rates for all cache levels.

\subsection{Setup}

Our target is a scalable ARM-based \gls{mpsoc} platform.
We use the \texttt{O3CPU} to model the ARM cores and Ruby for the cache sub-system.
To make our results comparable, we use the ARM CHI configuration provided in gem5 as the base.
Table~\ref{tab:target_system} shows its most relevant features.
The coherence protocol is ARM AMBA CHI~\cite{arm_amba_chi}, and the interconnect follows the hierarchical star topology shown in Fig.~\ref{fig:ruby-topo}.
We partition the system as described in Section~\ref{subsec:methods_partitioning}.

\begin{table}
  \centering
  \caption{Main Characteristics of the Simulated System}
  \begin{tabular}{|l|l|l|}
      \hline
    Component & Property & Value \\
    \hline
    \multirow{2}{*}{CPU} & Architecture & ARMv8-A 64-bit \\
      & Clock & 2GHz \\
    \hline
    \multirow{3}{*}{L1 I-Cache}
      & Capacity & 32 KiB \\
      & Associativity & 2 \\
      & Access latency & 1ns \\
    \hline
    \multirow{3}{*}{L1 D-Cache}
      & Capacity & 64 KiB \\
      & Associativity & 2 \\
      & Access latency & 1 ns \\
    \hline
    \multirow{3}{*}{L2 Cache}
      & Capacity & 2 MiB \\
      & Associativity & 8 \\
      & Access latency & 4 ns \\
      \hline
      \multirow{3}{*}{L3 Cache}
      & Capacity & 16 MiB \\
      & Associativity & 8 \\
      & Access latency & 6 ns \\
    \hline
    \multirow{2}{*}{DRAM} & Clock & 1GHz \\
    & Capacity & 512 MiB \\
    \hline
    \multirow{2}{*}{NoC} & Link and router latency & 0.5ns \\
    & Router buffer size & 4 messages (32-bit each) \\
    \hline
  \end{tabular}
  \label{tab:target_system}
\end{table}

This latency of links, routers and cache accesses is considered for defining a meaningful range for the simulation quantum.
In our configuration, travelling from the L1 cache router to the L2 cache router and then to the L3 cache router and returning through the same path involves crossing ten links for a total of 5ns.
If we add the cache access latencies from Table~\ref{tab:target_system}, we obtain an L3 cache hit latency of 16ns.
We set this value as the maximum quantum $t_q{\Delta}$ to bound the link latency artefacts from the inter-domain scheduling to a reasonable range.

We selected the following SW workloads for our evaluation:
A custom synthetic bare-metal benchmark, a sub-set of PARSEC v3.0\footnote{\url{https://parsec.cs.princeton.edu/download.htm}} and STREAM\footnote{\url{https://www.cs.virginia.edu/stream/}}.

The synthetic benchmark consists of a bare-metal multi-core test program designed to maximise CPU core utilisation while keeping the memory traffic low.
Each CPU executes a sorting algorithm on an exclusive memory region.
The loop and the data array are kept small so all instructions and data fit within a core's private caches.
There is no data sharing, and the input size is scaled linearly with the number of cores.

For PARSEC, we ran the applications shown in Table~\ref{tab:parsec_apps} with the \texttt{simmedium} input size setting.
The characteristics shown in the table are taken from the original author~\cite{parsec_summary_phd}.
Additionally, we ran STREAM, a benchmark used to show multi-core systems' maximum achievable DDR bandwidth.
PARSEC and STREAM run on top of the Ubuntu Linux 14.04 ﬁle system released by gem5\footnote{\url{https://www.gem5.org/documentation/general_docs/fullsystem/guest_binaries}}.

Our host platform is powered by an AMD Ryzen 3990x (64 dual-thread x86-64 cores),
has 128GiB of 3200MHz DDR4-DRAM and runs Ubuntu Linux 20.04.
For the compilation of PARSEC v3.0 for ARM 64-bit, we followed the official documentation from ARM\footnote{\url{https://github.com/arm-university/arm-gem5-rsk}}.
We employ fast-forwarding with the \texttt{AtomicCPU} and the checkpointing mechanism to simulate only the \glsfirstplural{roi} with the detailed timing mode.

\begin{table}\label{tab:parsec_apps}
  \centering
  \caption{PARSEC Application Characteristics}
    \begin{tabular}{|c|c|c|c|c|c|c|}
        \hline
        \multirow{2}{*}{Program} & \multicolumn{2}{c|}{Parallelisation} & \multicolumn{2}{c|}{Data Usage} \\
        \cline{2-5}
                                 & Model & Granularity                 & Sharing & Exchange \\
        \hline
        \texttt{blackscholes} & data-parallel & coarse & low  & low\\
        \texttt{canneal}      & unstructured  & fine   & high & high\\
        \texttt{dedup}        & pipeline      & medium & high & high\\
        \texttt{ferret}       & pipeline      & medium & high & high\\
        \texttt{fluidanimate} & data-parallel & fine   & low  & medium\\
        \texttt{swaptions}    & data-parallel & coarse & low  & low\\
        \hline
    \end{tabular}
\end{table}

\begin{figure}[bt]
  \centering
  \includegraphics[width=0.99\linewidth]{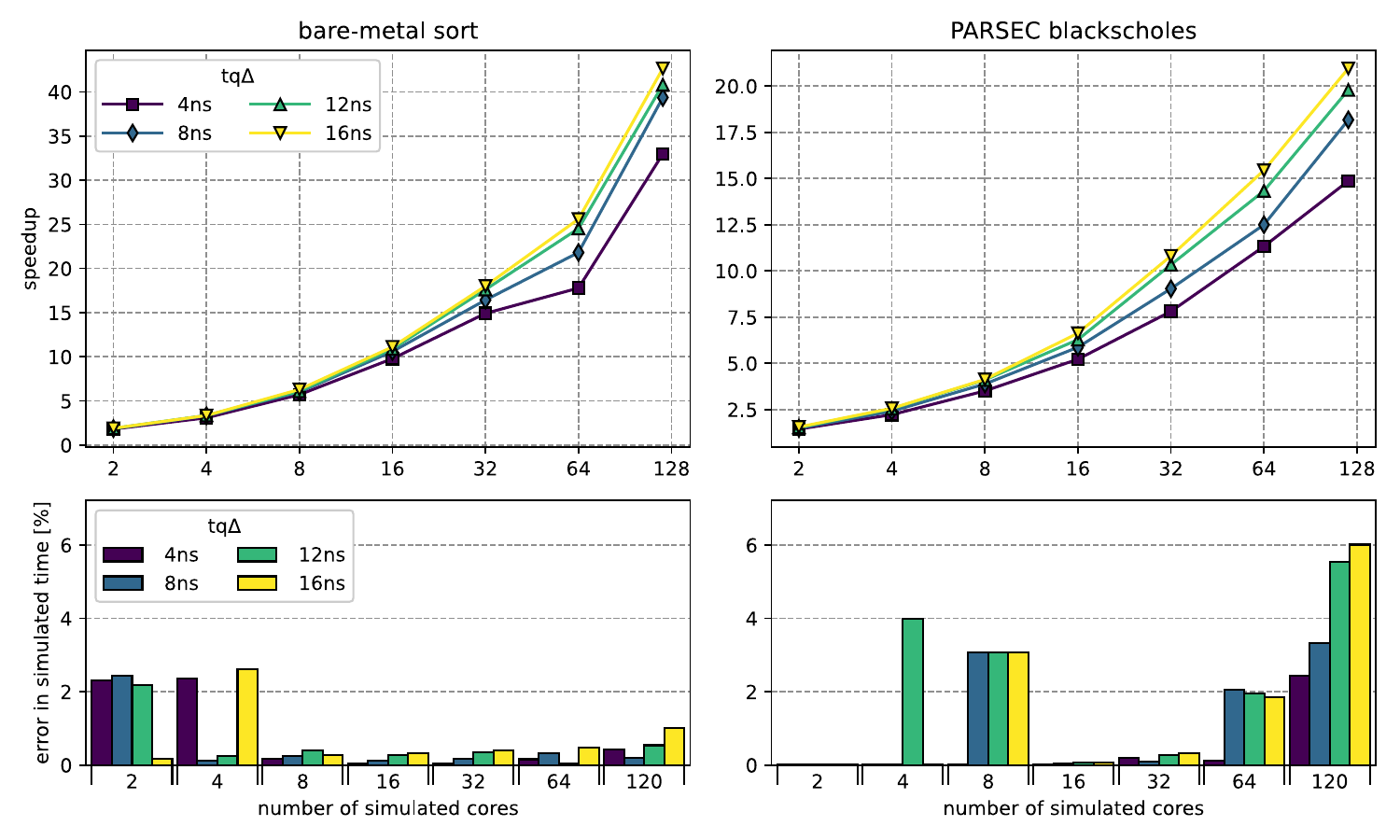}
  \caption{Speedup and simulation error for our bare-metal application (left) and PARSEC \texttt{blackscholes} (right),
           as a function of the number of simulated cores and quantum setting.}
  \label{fig:bm_speedup}
\end{figure}

\subsection{Results}

\subsubsection{Speedup and Simulated Time Accuracy}

As a starting point, we perform a core and quantum sweep with our synthetic benchmark and PARSEC's \texttt{blackscholes}.
The number of cores is increased in multiples of two, stopping at 120 cores so as not to create more software threads than are physically available.

The resulting speedups can be observed in the top plots in Fig.~\ref{fig:bm_speedup}.
Both applications scale well with the core count.
Though, as expected, the bare-metal application exhibits the highest speedups.
The bare-metal program achieves a maximum observed speedup of 42.7$\times$ when simulating 120 cores.
The error in simulated time is below 3\% in all cases.
For \texttt{blackscholes}, the maximum speedup is 21.0$\times$, and the error grows to 6.0\% for the largest quantum setting.

\begin{figure}[bt]
  \centering
  \includegraphics[width=\linewidth]{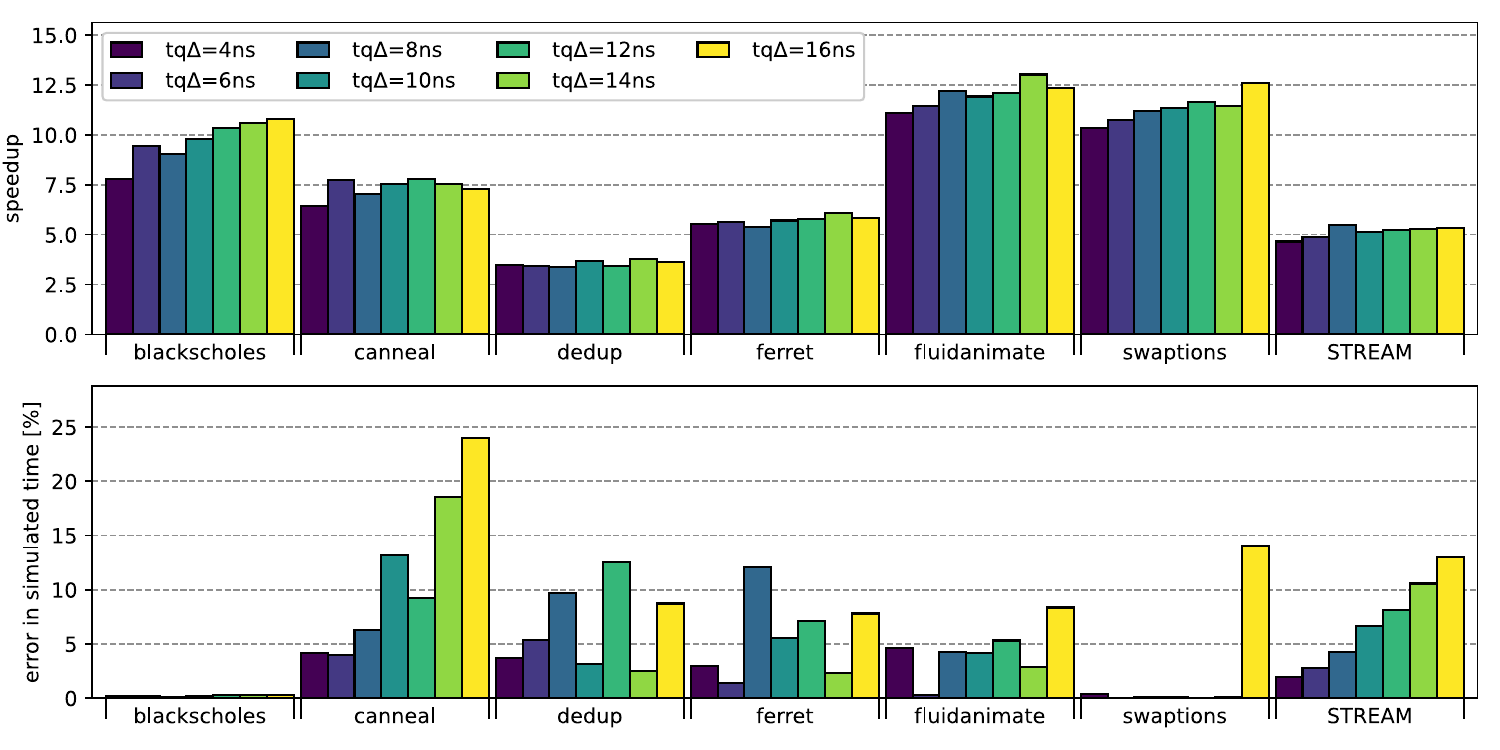}
  \caption{Speedup and simulation error of PARSEC and STREAM on a 32-core target system. The colour of the bars indicates the quantum setting.}
  \label{fig:parsec_speedup}
\end{figure}

We fixed the hardware platform to 32 cores to evaluate the performance and accuracy for the remaining multi-thread benchmarks.
The resulting speedup and the error in total simulated time is shown in Fig.~\ref{fig:parsec_speedup}.
The application \texttt{swaptions} yields the highest performance gain with a remarkable speedup of 12.6$\times$.
As a result, the simulation, typically requiring over a full day, is completed in approximately 2.3 hours.
On the other end, \texttt{dedup} achieves only 3.6$\times$. The average speedup is 10.7$\times$.

The variation in speedup among applications indicates that the simulated workload's data access patterns strongly influence the achievable acceleration.
The error in simulation also displays a high dependency on the application.
Moreover, it can be noted that the test programs characterised by high data sharing and high data exchange yield the lowest speedup and the highest error.
This is the case of \texttt{canneal}, \texttt{dedup} and \texttt{ferret}, as specified in Table~\ref{tab:parsec_apps}.
The reason is simple: if the amount of shared data is high, cache conflicts and accesses to the main memory will occur more often.
This harms the performance since accesses to shared resources are serialised with mutexes.
At the same time, a high number of cross-domain events results in more timing deviations.
STREAM, designed to generate as many accesses to the off-chip memory as possible, also falls in this category.
This observation is also made in \cite{sniper2011}.

The quantum also influences performance and accuracy.
Although the trend is not so pronounced for all test programs, small quanta are needed to keep the error low.
Setting the quantum to 12ns or less, leads to simulation errors below 15\% for all considered applications.
Such a constraint also limits the performance, but the loss in speedup is only between 1\% and 8\%.

\subsubsection{Cache Hit Rates}

\begin{figure*}[bt]
  \centering
  \includegraphics[width=\textwidth]{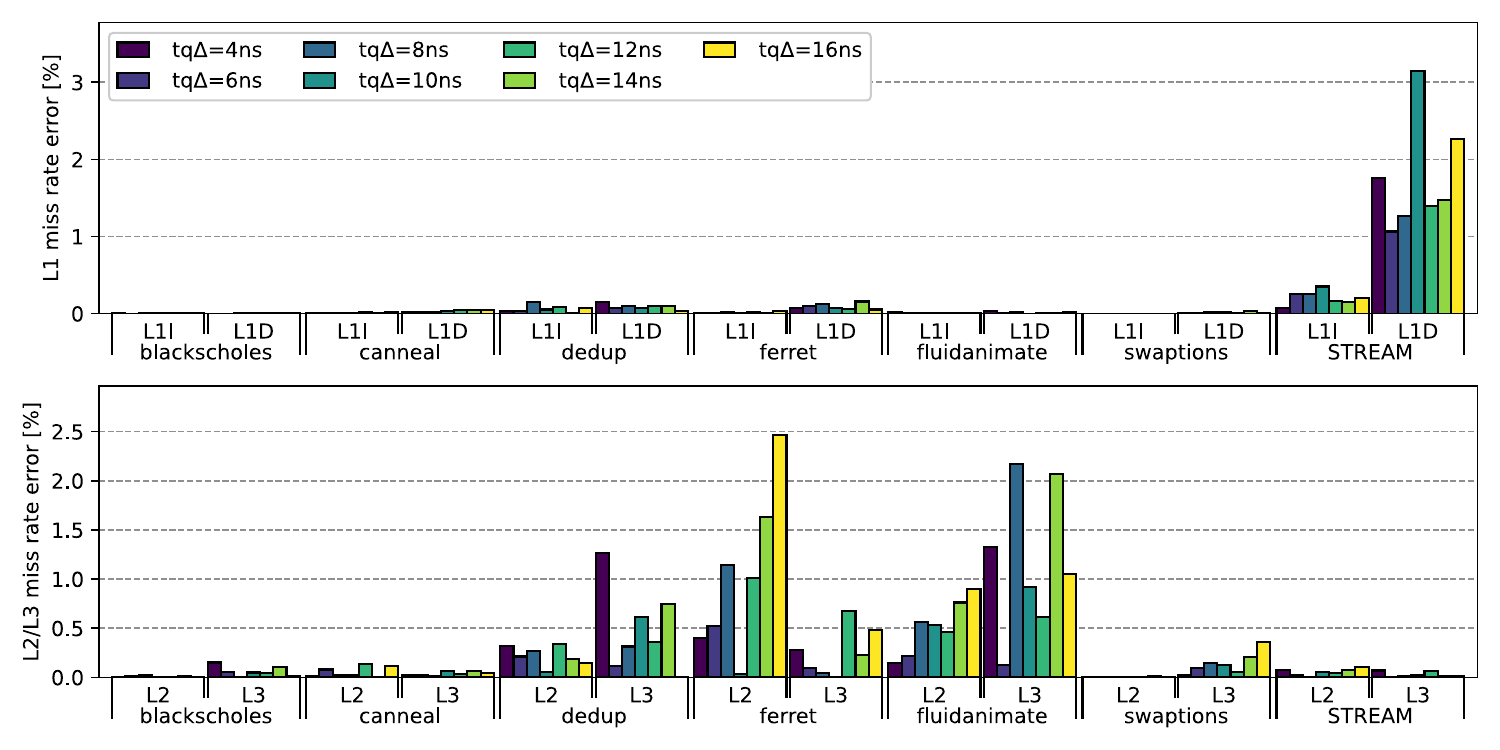}
  \caption{Error on cache miss rates for PARSEC and STREAM on a 32-core target system.}
  \label{fig:parsec_cache}
\end{figure*}

We look into the cache statistics to further evaluate the accuracy of \emph{parti-gem5} simulations.
In Fig.~\ref{fig:parsec_cache}, we plot the absolute error in the miss rate for all cache levels.
For private L1 and L2 caches, we compute the average among all cores.
The absolute error for this vital metric remains below 2.5\% for all applications and quantum values.

\section{Conclusion \& Outlook}\label{sec:conclusion}

Our work demonstrates that the acceleration of detailed timing simulations for \glspl{mpsoc} on gem5 is achievable through the use of \gls{pdes},
effectively harnessing the computational power of modern multi-core simulation hosts.
In this context, \emph{parti-gem5} serves as a valuable addition to existing methods, such as sampling and checkpointing,
by enabling rapid exploration of micro-architectures.
One notable advantage of our approach is its ability to simulate larger portions of the target software applications with detailed models.
The extent of speedup achieved relies on the scalability of the simulated multi-thread software workload.
Our evaluations reveal that applications based on barriers and those with limited data sharing derive the greatest benefit from \emph{parti-gem5}.

Despite the introduction of timing inaccuracies due to the non-determinism of parallel simulations,
our findings demonstrate that the deviations in simulated time can be kept below 15\% without a significant sacrifice in throughput.
To achieve this, it is crucial to set meaningful quantum values based on the latencies of the target system.
Although we did not encounter any causality errors that affected the correctness of the simulated workloads,
conducting more formal verification to validate the correctness and preservation of memory consistency would be an important contribution to our work.

While our evaluation was focused on the ARM \gls{isa}, the concepts we introduced can be extended to any other target architecture.
Further experimentation is necessary to explore simulation performance for other interconnect topologies and software workloads.
Nevertheless, we hope our contributions inspire and enable further advancements in this field.


\end{document}